\documentclass[prb,showpacs,floatfix,aps,twocolumn]{revtex4}
\usepackage{epsfig,amsmath,amssymb}
\newcommand{\be}{\begin{equation}}
\newcommand{\ee}{\end{equation}}

\newcommand{\bit}{\begin{itemize}}
\newcommand{\eit}{\end{itemize}}
\newcommand{\bea}{\begin{eqnarray}}
\newcommand{\eea}{\end{eqnarray}}
\newcommand{\Kagome}{{kagom\'e} }
\begin{document}
\title
{
Green's function theory of quasi-two-dimensional spin-half Heisenberg 
ferromagnets: stacked square 
 versus stacked \Kagome lattice
}
\author
{
D. Schmalfu{\ss}$^{a}$, J. Richter$^{a}$ and D. Ihle$^{b}$ 
}
\affiliation
{
$^{a}$Institut f\"ur Theoretische Physik, Universit\"at
Magdeburg, 39016 Magdeburg, Germany \\   
}
\affiliation
{
$^{b}$Institut f\"ur Theoretische Physik,Universit\"at Leipzig, 04109 Leipzig,
Germany\\
}                     

\begin{abstract}
We consider the thermodynamic properties of the  quasi-two-dimensional 
spin-half Heisenberg ferromagnet on the stacked square and the stacked
\Kagome lattices by using the
spin-rotation-invariant Green's function method.  We 
calculate the critical temperature $T_C$,
the uniform static susceptibility $\chi$, the correlation
lengths $\xi_\nu$ and  the magnetization $M$
and investigate the short-range order above $T_C$.
We find that $T_C$ and $M$ at $T>0$
 are smaller for the  stacked
\Kagome lattice  which we attribute to frustration effects becoming
relevant at finite temperatures.
\end{abstract}

\pacs{
75.10.Jm;	
75.45.+j;	
75.50.Ee	
}

\maketitle

{\it Introduction:}
Quasi-two-dimensional magnets have attracted much attention
in recent years.\cite{wir04,lhuillier03}
 According to the Mermin-Wagner theorem\cite{merm} 
strictly 
two-dimensional (2D) Heisenberg magnets do not possess magnetic long-range order
(LRO) at any finite temperature. Therefore, the exchange coupling between
planes is crucial for the existence of a finite critical temperature
$T_C$.
Though most of the quasi-2D magnetic insulators 
are antiferromagnets there are also some quasi-2D ferromagnetic insulators 
like 
$K_2CuF_4$, $La_2BaCuO_5$, $Cs_2AgF_4$.\cite{feld95} 
The calculation of thermodynamic properties for quasi-2D
magnets is an important issue,\cite{irkhin0,kopietz,ishi93,
irkhin1,irkhin2} in particular with respect to the
interpretation of  experimental results. 
Especially in the limit of weak interlayer coupling the evaluation of $T_C$
is challenging.  
Recently
accurate values for 
the critical temperature of a
Heisenberg magnet on a stacked square
lattice by 
Monte-Carlo calculations\cite{yasuda}   and 
on the cubic lattice by 
high-temperature series expansion\cite{oitmaa}  
have been  presented. 
Another promising approach to calculate the thermodynamics of
quasi-2D magnetic systems 
is a spin-rotation
invariant Green's function method (RGM).\cite{kondo,shima,BB94,winter,siu,bern,schmal,junger,phd}
The RGM allows a consistent description of LRO as well as
of short-range order (SRO) at arbitrary temperatures in magnetic systems 
of arbitrary dimension. 

In this paper we use the RGM
for the calculation of thermodynamic quantities of
the quasi-2D spin-half Heisenberg 
ferromagnet on the
stacked square and the
stacked \Kagome lattice. Both lattices have the same coordination
number, but differ in their lattice geometry within the layers. 
Note that corresponding  RGM treatments 
for the respective stacked antiferromagnets were given in 
Refs.~\onlinecite{siu} and \onlinecite{schmal}. 
The Heisenberg
antiferromagnet on the \Kagome
lattice is a canonical model to study frustration effects in 2D
spin-half systems, see, e.g. 
Refs.~\onlinecite{bern,lecheminant97,waldtmann98,jump,wir04}. 
Due to strong frustration  there is most likely 
no magnetic LRO in
the \Kagome  antiferromagnet.
It has been argued recently\cite{schmal} 
that even the increase of the dimension in 
a stacked \Kagome
antiferromagnet does not lead to magnetic LRO.

In this paper we consider layered  ferromagnets.
Then geometric frustration is
irrelevant in the ground state. However, at finite temperatures
excited states with antiferromagnetic correlations contribute to the
partition function. The energy of such states is influenced by
triangular spin configurations  and thus the frustrating
geometry of the \Kagome lattice may become more and more relevant with
increasing temperature. By comparison with the
unfrustrated stacked square 
lattice we can discuss these frustration effects.
\begin{figure}
\begin{center}
\epsfig{file=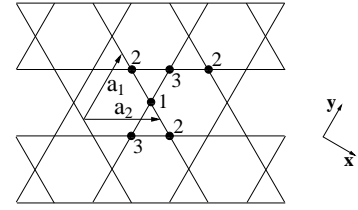,scale=0.255,angle=0.0}
\end{center}
\caption{ 
Sketch of one  \Kagome layer in  the congruently stacked
lattice with the in-plane geometric
basis vectors
${\bf{a}}_{1}=\left(0,2,0\right),{\bf{a}}_{2}=\left(\sqrt{3},1,0\right)$.
The out-of-plane basis vector
${\bf{a}}_{3}=\left(0,0,1\right)$ is not shown.
The geometrical unit cell contains three sites (spins)
labeled by a running index $\alpha=1,2,3$.}
\label{fig1}
\end{figure}

We consider  the Heisenberg model of $N$ spins $1/2$
\be
H=\frac
{1}{2}\sum_{m\alpha,n\beta}
J_{m\alpha,n\beta}{\bf{S}}_{m\alpha}{\bf{S}}_{n\beta}, \quad 
J_{m\alpha,n\beta} \le 0,
\label{eq1}
\ee
where the sum runs over all $\mathcal{N}$ unit cells (labeled by $m$ and
$n$) and all spins within a unit cell (labeled by running indices $\alpha$ and
$\beta$, see Fig. \ref{fig1} for the \Kagome case). 
The exchange coupling $J_{m\alpha,n\beta}$ is nonzero for nearest
neighbors, only. The  ferromagnetic  exchange parameter within
the layers $J_{\parallel}$ is fixed to $J_{\parallel}=-1$.

{\it Spin-Rotation-Invariant Green's Function Method:} 
The rotation-invariant
decoupling scheme introduced by 
Kondo and Yamaji\cite{kondo} goes one step 
beyond the  random-phase approximation (RPA).
An advantage
of this method is the possibility to treat magnetic order-disorder
transitions driven by quantum fluctuations as well as by thermal fluctuations
and to provide a good description of magnetic SRO.\cite{kondo,shima,BB94,winter,siu,bern,schmal,junger,phd}
In this paper we
only illustrate some basic features of the RGM. For more details 
the reader is referred to
Refs.~\onlinecite{kondo,shima,BB94,winter,bern,junger}, and in particlar to
Refs.~\onlinecite{siu, schmal,phd} where the corresponding stacked
antiferomagnets are considered.

To evaluate all relevant correlation functions
$\langle {\bf{S}}_{m\alpha}{\bf{S}}_{n\beta}\rangle$ 
 and several thermodynamic quantities
we have to
calculate a set of Fourier-transformed commutator Green's functions
$\langle\langle
S^{+}_{{\bf{q}}\alpha};S^{-}_{{\bf{q}}\beta}\rangle\rangle_{\omega}$
which are connected with the dynamic spin susceptibilities
by
$\chi^{+-}_{{\bf{q}}\alpha\beta}(\omega)=-\langle\langle
S^{+}_{{\bf{q}}\alpha};S^{-}_{{\bf{q}}\beta}\rangle\rangle_{\omega}$.
Using the equations of motion and supposing  
spin-rotational invariance, i.e. $\langle S^{z}_{m\alpha}\rangle=0$,
we obtain
$
\omega^{2}\langle\langle
S^{+}_{{\bf{q}}\alpha};S^{-}_{{\bf{q}}\beta}\rangle\rangle_{\omega}
=\langle [i\dot{S}^{+}_{{\bf{q}}\alpha},S^{-}_{{\bf{q}}\beta}]\rangle
+\langle\langle
-\ddot{S}^{+}_{{\bf{q}}\alpha};S^{-}_{{\bf{q}}\beta}\rangle\rangle_{\omega}.
$
The operator $\; -\ddot{S}^{+}_{{\bf{q}}\alpha}=[[S^{+}_{{\bf
q}\alpha},H],H]$ written in site representation
contains products of three
spin operators along nearest-neighbor sequences which are 
treated in the spirit of the decoupling scheme of Shimahara and 
Takada.\cite{shima} 
For example, the product
$S^{-}_{A}S^{+}_{B}S^{+}_{C}$
is replaced by $\eta_{A,B}\langle S^{-}_{A}S^{+}_{B}\rangle
S^{+}_{C}+\eta_{A,C}\langle S^{-}_{A}S^{+}_{C}\rangle
S^{+}_{B}$, where $A,B,C$ represent spin sites. The  vertex parameters
$\eta_{\gamma,\mu}$  are introduced to improve the
approximation.
In the
minimal version of the RGM we introduce just as many vertex parameters as
independent conditions for them can be formulated, i.e. we have 
$\eta_{\parallel}$ and
$\eta_{\perp}$ attached to intralayer and interlayer correlators,
respectively. 
After the decoupling
we can write the equation of motion in a compact matrix form
omitting the running indices $\alpha$ and $\beta$,
$
\left(\omega^{2} -
F_{{\bf{q}}}\right)\chi^{+-}_{{\bf{q}}}\left(\omega\right)=
- M_{{\bf{q}}}$,
where $F_{{\bf{q}}}$ and $M_{{\bf{q}}}$ are the frequency and momentum
matrices, respectively. Their dimension equals 
the number of spins in the respective unit cell. The
solution for $\chi^{+-}_{{\bf{q}}}\left(\omega\right)$
can be formulated in terms of the common set of normalized
eigenvectors $\left|j{\bf{q}}\right\rangle$ of $F_{{\bf{q}}}$ 
and $M_{{\bf{q}}}$ 
and reads
$
\chi^{+-}_{{\bf{q}}}\left(\omega\right)=-\sum_{j=1}^{n}\frac
{m_{j{\bf{q}}}}{\omega^{2}-\omega^{2}_{j{\bf{q}}}}\left|j{\bf{q}}\right\rangle\left\langle
j{\bf{q}}\right|$.
The case $J_{\parallel}>0$ for the stacked \Kagome lattice has been discussed 
previously \cite{schmal}
in detail. Hence, it is not necessary to present the lengthy expressions for the eigenvalues $m_{j{\bf{q}}}$
of $M_{{\bf{q}}}$ and $\omega^{2}_{j{\bf{q}}}$ of $F_{{\bf{q}}}$ once again. 
The corresponding expressions for $m_{j{\bf{q}}}$ and
$\omega^{2}_{j{\bf{q}}}$ in case of 
the stacked square lattice are even simpler,
since  only one
eigenvector $\left|1{\bf{q}}\right\rangle$ exists. 
For both lattices 
the eigenvalues $m_{j{\bf{q}}}$
and $\omega^{2}_{j{\bf{q}}}$ contain  eight parameters which must be
determined selfconsistently, namely the spin-spin
correlators
$c_{1,0,0},c_{1,1,0},c_{2,0,0},c_{0,0,1},c_{1,0,1},c_{0,0,2}$ and the two
vertex parameters $\eta_{\parallel}$,
and $\eta_{\perp}$.   
The correlators are defined by $c_{k,l,m}\equiv c_{{\bf{R}}}=\left\langle
S^{+}_{{\bf{0}}}S^{-}_{{\bf{R}}}\right\rangle=2\left\langle
{\bf{S}}_{{\bf{0}}}{\bf{S}}_{{\bf{R}}}\right\rangle/3$, where 
${\bf{R}}=k{\bf{b}}_{1}+l{\bf{b}}_{2}+m{\bf{b}}_{3}$ 
(the ${\bf b}_{i}$ 
are the basis vectors of the cubic lattice)
for the stacked square lattice, but 
${\bf{R}}=k{\bf{a}}_{1}/2+l{\bf{a}}_{2}/2+m{\bf{a}}_{3}$ for the stacked
\Kagome lattice, cf.  Fig.~\ref{fig1}.
The spectral
theorem\cite{tya} for the correlators $c_{k,l,m}$ yields seven equations including the
sum rule $c_{0,0,0}=1/2$. According to  
Refs.~\onlinecite{siu,schmal} and \onlinecite{phd} an eighth equation is obtained requiring that the
matrix of the static susceptibility $\chi^{+-}_{{\bf{q}}}=\chi^{+-}_{{\bf{q}}}\left(\omega=0\right)$ has to be isotropic in the limit
${\bf{q}}\to{\bf{0}}$, i.e., $\lim_{q_{x\left(y\right)}\to 0}\left.\chi^{+-}_{{\bf{q}}}\right|_{q_{z}=0}=\lim_{q_{z}\to
0}\left.\chi^{+-}_{{\bf{q}}}\right|_{q_{x\left(y\right)}=0}$.
In the RGM scheme magnetic LRO is detected 
by\cite{shima,BB94,winter,siu,junger} 
$\lim_{\left|{\bf{R}}\right|\to\infty}\left\langle
{\bf{S}}_{{\bf{0}}}{\bf{S}}_{{\bf{R}}}\right\rangle \ne 0$, i.e. the
correlation function contains 
a long-range part $C_{{\bf{Q}}\alpha\beta}$, introduced by 
Shimahara and  Takada 
and called condensation part\cite{shima}, leading to 
$
\left\langle
{\bf{S}}_{m\alpha}{\bf{S}}_{n\beta}\right\rangle=\frac
{1}{\mathcal{N}}\sum_{{\bf{q}}\neq{\bf{Q}}}S_{{\bf{q}}\alpha\beta}
e^{-i{\bf{q}}{\bf{r}}_{m\alpha,n\beta}}
+\frac {3}{2}C_{{\bf{Q}}\alpha\beta}e^{-i{\bf{Q}}{\bf{r}}_{m\alpha,n\beta}},
$
where
${\bf{Q}}$ is the magnetic wave vector and
$
S_{{\bf{q}}\alpha\beta}=\frac {3}{2}\sum_{j=1}^{n}\frac
{m_{j{\bf{q}}}}{2\omega_{j{\bf{q}}}}\left(1+\frac{2}{e^{\omega_{j{\bf{q}}}/T}
-1}\right)\left\langle\alpha\right.
\left|j{\bf{q}}\right\rangle\left\langle
j{\bf{q}}\right.\left|\beta\right\rangle$.
The existence of  nonzero $C_{{\bf{Q}}\alpha\beta}$
is
accompanied by a diverging static susceptibility $\chi^{+-}_{{\bf{q}}}$ at
 ${\bf{q}}={\bf{Q}}$. This gives an additional equation for
one $C_{{\bf{Q}}\alpha\beta}$. If the  unit cell contains more
than one spin, the further 
relations for the various condensation terms 
corresponding  to different Green's functions  are given by
\cite{phd}
$
C_{{\bf{Q}}\alpha\beta}/C_{{\bf{Q}}\gamma\delta}=
\lim_{{\bf{q}}\to{\bf{Q}}}\chi^{+-}_{{\bf{q}}\alpha\beta}/\chi^{+-}_{{\bf{q}}\gamma\delta}.
$
The condensation term $C_{{\bf{Q}}\alpha\beta}$ is related to the order
parameter $M_{\alpha\beta}$ by
$M^{2}_{\alpha\beta}=3\left|C_{{\bf{Q}}\alpha\beta}\right|/2$ with
$M^{2}_{\alpha\alpha}$ being the square of the 
magnetization \cite{shima} of a certain site $\alpha$ in the unit cell.
In a short-range ordered phase the condensation terms are zero.
For temperatures lower than the respective Curie temperature $T_C$
both systems exhibit ferromagnetic LRO with 
${\bf{Q}}=\left(0,0,0\right)$. 
In the
\Kagome case we have three spins per unit cell, and in principle six
nonequivalent Green's functions have to be considered. However, due to 
symmetry we 
have 
$C_{{\bf{Q}}\alpha\beta}/C_{{\bf{Q}}\gamma\delta}=1$, i.e. all 
three sites in the unit cell carry the same magnetization
$M=M_{\alpha\alpha}$, $\alpha=1,2,3$.
The condensation term 
is obtained by demanding that 
the static susceptibility in the long-range ordered phase
diverges at the magnetic wave vector ${\bf{Q}}$, which yields
$
J_{\parallel}\left(1-2\eta_{\parallel}\left(4c_{1,0,0}-c_{1,1,0}-c_{2,0,0}\right)\right)
+4J_{\perp}\left(\eta_{\perp}c_{1,0,1}-\eta_{\parallel}c_{1,0,0}\right)=0 
$
for the stacked \Kagome lattice and 
$
J_{\parallel}\left(1-2\eta_{\parallel}\left(5c_{1,0,0}-2c_{1,1,0}-c_{2,0,0}\right)\right)
+4J_{\perp}\left(\eta_{\perp}c_{1,0,1}-\eta_{\parallel}c_{1,0,0}\right)=0
$
for the stacked square lattice.  At zero temperature
the solution of our set of equations becomes exact, namely we find
$c_{k,l,m}=C_{{\bf{Q}}\alpha\beta}=1/6$ and
$\eta_{\parallel}=\eta_{\perp}=3/2$, and consequently 
$\left\langle
{\bf{S}}_{{\bf{0}}}{\bf{S}}_{{\bf{R}}}\right\rangle=1/4$ 
and $M_{\alpha\alpha}=1/2$. 

Further quantities of interest are 
the uniform static susceptibility 
$\chi=\lim_{{\bf{q}}\to{\bf{0}}}\chi^{+-}_{{\bf{q}}}/2$ and the 
 intralayer and interlayer correlation lengths
$\xi_{\parallel}$ and $\xi_{\perp}$, which 
can be calculated
as described in  Refs.~\onlinecite{shima,winter,siu,schmal,phd}.

\begin{figure}
\begin{center}
\epsfig{file=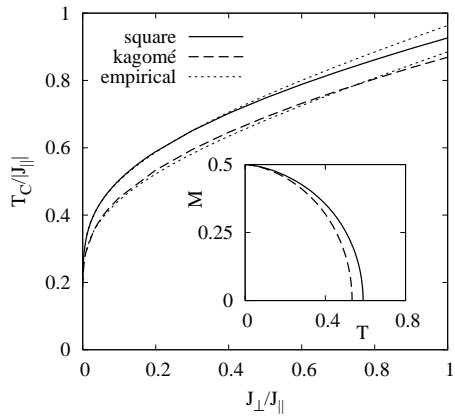,scale=0.315,angle=270.0}
\end{center}
\caption{Curie temperature $T_{C}$ as a function of the interlayer
coupling $J_{\perp}/J_{\parallel}$. For comparison we give the
empirical curves according to Eq.~(\ref{empi}).
The inset shows the magnetization $M$ in dependence on temperature $T$ for
$J_{\perp}/J_{\parallel}=0.2$.}
\label{fig_tcrit}
\end{figure}

{\it Results:} 
The critical temperature  in dependence on the interlayer coupling $J_\perp$
is plotted in Fig.~\ref{fig_tcrit}. 
Note that  the 
RGM fulfills
the Mermin-Wagner theorem \cite{merm}, i.e. $T_C(J_\perp =0)=0$.
In the inset of Fig.~\ref{fig_tcrit} we show $M(T)$ 
curves for 
$J_\perp=-0.2$. Obviously, 
the magnetization $M$ at $0 < T < T_C$ and the value of $T_C$ for the stacked \Kagome
lattice are always smaller than $M$ and $T_C$  for the stacked square 
lattice. 
The relation $T_C^{kagome} < T_C^{square}$ holds for any
value of $J_\perp$ (see Fig.~\ref{fig_tcrit}). To give a possible
interpretation for that  
we have to look at the energy spectrum for the strictly 
2D square and \Kagome
lattices. The ground-state energy (lower bound of the spectrum)
is $E_{min}=-0.5N$  for both lattices. 
However, 
the upper bound  for the \Kagome lattice $E_{max}=0.434N$ is 
much lower than the
corresponding value $E_{max}=0.670N$ for the square lattice 
(see, e.g., Ref.~\onlinecite{wir04}). This upper bound is equal to the 
ground-state energy of the  \Kagome {\it antiferromagnet} 
(i.e. $J_\parallel=+1$) multiplied by minus one, and its small 
value signals
strong frustration effects being relevant for eigenstates with 
antiferromagnetic spin-spin correlations. 
Hence, one can expect that
excited states with antiferromagnetic spin correlations have lower energy
for the \Kagome ferromagnet resulting in a 
larger contribution to the partition function at $T>0$ in comparison with 
the square-lattice ferromagnet.
The observation $T_C^{kagome} < T_C^{square}$ 
corresponds to the data for the 
susceptibility $\chi$ and the correlation lengths $\xi_\nu$, 
i.e., $\chi$ and  $\xi_\nu$
for the  stacked \Kagome lattice 
are smaller than the corresponding values  
for the stacked square
lattice (see below, Fig.~\ref{fig_xi}). 
\begin{figure}
\begin{center}
\epsfig{file=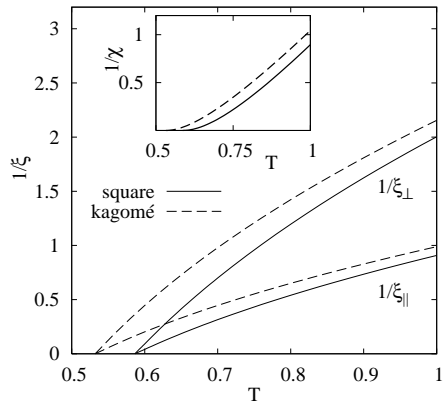,scale=0.315,angle=270.0}
\end{center}
\caption{Inverse intralayer ($\xi_{\parallel}^{-1}$) and interlayer
($\xi_{\perp}^{-1}$) correlation lengths 
vs. $T$
for $J_{\perp}/J_{\parallel}=0.2$. The inset shows the 
inverse uniform static susceptibility $\chi$ 
for $J_{\perp}/J_{\parallel}=0.2$.}
\label{fig_xi}
\end{figure}
\begin{table}
\begin{ruledtabular}
\begin{tabular}{ccccc}
$J_{\perp}/J_{\parallel}$  & square FM  & \Kagome FM & square AFM    & square AFM \\
                           & RGM        & RGM      & Schwinger\cite{irkhin0}    & QMC\cite{yasuda}    \\ \hline
$1$     & $1.2351$ & $1.1588$  & $1.51 $ & $1.2589 $\\
$0.5$   & $0.9967$ & $0.9213$  & $1.24 $ & $1.0050 $\\
$0.1$   & $0.6723$ & $0.6029$  & $0.91 $ & $0.6553 $\\
$0.05$  & $0.5851$ & $0.5209$  & $-  $   & $0.5689 $\\
$0.01$  & $0.4439$ & $0.3917$  & $0.63$  & $0.4352$ \\
$0.005$ & $0.4004$ & $0.3556$  & $-  $   & $0.3951 $\\
$0.001$ & $0.3244$ & $0.2895$  & $ - $   & $0.3257 $\\
\end{tabular}    
\end{ruledtabular}
\caption{\label{tab_1}Curie temperatures
$T_{C}/\left(|J_{\parallel}|s\left(s+1\right)\right)$ of the $s=1/2$ ferromagnet
on the stacked square and stacked kagom\'{e} lattice calculated by the RGM.
For comparison we show some results 
for the Ne\'el temperature $T_N/\left(J_{\parallel}s(s+1)\right)$ of 
the stacked square-lattice antiferromagnet obtained by other approaches.
For the simple cubic lattice 
high-order 
high-temperature series expansion\cite{oitmaa} yields 
 $T_{C}/\left(|J_{\parallel}|s\left(s+1\right)\right)=1.119$ and 
 $T_{N}/\left(|J_{\parallel}|s\left(s+1\right)\right)=1.259$.}
\end{table}

In the limit $J_{\perp}/J_{\parallel}\ll 1$ we derive
more explicit expressions 
for $T_{C}$ following Ref.~\onlinecite{kopietz}. We find
$
T_{C}=-J_{\parallel}\pi/(2\ln(-20T_{C}/9J_{\perp}))
$
for the stacked square lattice and
$
T_{C}=-3J_{\parallel}\pi/(8\ln(-20T_{C}/9J_{\perp}))
$
for the stacked \Kagome lattice. Note that these expressions are of 
similar form as corresponding 
formulas for the 
Ne\'el temperature $T_N$ obtained by  selfconsistent spin-wave theories. 
\cite{kopietz,irkhin1,irkhin2}
We compare our data for $T_C$ in Tab.~\ref{tab_1} with
results for the Ne\'el temperature $T_N$ 
of the  
stacked square-lattice antiferromagnet. 
For the simple cubic lattice  (i.e. 
$J_\perp =J_\parallel$) our result for $T_C$ 
exceeds the best available value for $T_C$ obtained by high-order 
high-temperature series expansion\cite{oitmaa} by 9\% which can be considered
as a test of the accuracy of the RGM results. This accuracy is much
better than that of the Schwinger boson approach 
(compare 4th and 5th columns in
Tab.~\ref{tab_1}). Note that the accurate 
results of Ref.~\onlinecite{oitmaa} show that $T_N > T_C$
for  unfrustrated spin-half Heisenberg magnets, 
whereas  an RPA treatment\cite{du} yields 
$T_N=T_C$.  Therefore,  
the relation between $T_C$ and $T_N$ 
 can be used as a further  test of the RGM.
Indeed, using the
same set of vertex parameters\cite{vgl} 
 for  $J_\perp =
J_\parallel$ we obtain  $T_N= 1.1225 T_C$ which is in excellent agreement
with Ref.~\onlinecite{oitmaa}.

The overall dependence of $T_C$ versus $J_\perp$ resembles 
the results of
Yasuda et al.\cite{yasuda} for $T_N$, 
see Tab. \ref{tab_1}. 
Therefore, we adopt their empirical  formula 
\be \label{empi}
\frac{T_C}{|J_\parallel|} = \frac{A}{B-\ln(J_\perp/J_\parallel)}.
\ee
The best agreement with our data is obtained for $A=2.414 \; (2.049)$ and 
$B=2.506 \; (2.315)$  for the stacked square 
(kagom\'e) ferromagnet, cf. Fig.~\ref{fig_tcrit}. 

\begin{figure}
\begin{center}
\epsfig{file=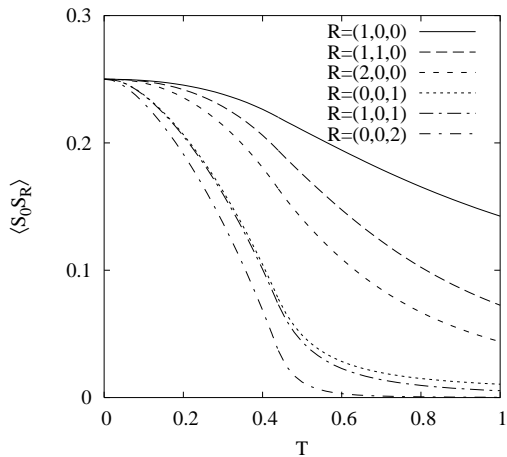,scale=0.35,angle=270.0}
\end{center}
\caption{Short-range intra- and interlayer correlation functions 
$\langle {\bf S}_0{\bf S}_{{\bf R}} \rangle$ for the stacked 
square
lattice with $J_{\perp}/J_{\parallel}=0.05$.  
}
\label{fig_sro}
\end{figure}

The correlation lengths and  
the uniform static susceptibility 
are shown as functions of T in Fig.~\ref{fig_xi} for 
$J_{\perp}/J_{\parallel}=0.2$.
As expected, those quantities  
diverge at $T_C$ and they go to zero for $T \to \infty$.
The inverse susceptibility depends   linearly on $T$ for temperatures
well above $T_C$, but it deviates from the linear Curie-Wei{\ss} 
law near  $T_C$ (see inset of Fig.~\ref{fig_xi}).

Finally, we discuss the short-range 
spin-spin correlators.
As an example we present their temperature dependence  
for the stacked 
square lattice for a weak interlayer coupling  $J_\perp= - 0.05$ 
in Fig.~\ref{fig_sro}.
For this value of $J_\perp$ we have $T_C=0.4388$, i.e. for $T > 0.4388$ the
long-range part of the spin-spin correlation function vanishes.
It is obvious  that the short-range 
 intralayer and interlayer correlators
behave differently. 
The interlayer correlators 
become very small at $T= T_C$
whereas the intralayer spin-spin-correlations are still 
well-pronounced.
This behavior corresponds to the data for the 
intra- and interlayer correlation
 lengths, see Fig.~\ref{fig_xi}, and
indicates the crossover from three-dimensional magnetic 
LRO at $T < T_C$ 
to 2D SRO at $T > T_C$, if $|J_\perp | \ll 
|J_\parallel |$.

{\it Summary:}
In this paper the thermodynamics of the spin-half Heisenberg
ferromagnet on the stacked square  and  \Kagome lattices has been 
investigated applying a 
spin-rotation-invariant Green's function method. 
We have calculated the Curie temperatures $T_{C}$   in
dependence on the interlayer coupling $J_\perp$, and simple empirical
formulas for $T_C(J_\perp)$
are obtained from the numerical data. Studying short-range
spin-spin correlation functions we see clearly a dimensional crossover
at $T \sim T_C$ for strongly anisotropic systems. 
Comparing the values of $T_{C}$, of the magnetization,
the correlation lengths and of 
the uniform static susceptibility we come to the
conclusion that
at $T>0$
frustration becomes relevant   
for the stacked \Kagome lattice
leading to a weakening of magnetic order at finite  
temperatures and to  $T_C^{kagome} < T_C^{square}$.

{\it   Acknowledgment:}
This work was supported by the DFG (project Ri615/12-1, Ih13/7-1).


\begin{thebibliography}{100}
\bibitem{vgl} The stacked square-lattice Heisenberg antiferromagnet
was treated within the RGM in Ref.~\onlinecite{siu}. However, 
in that paper an additional vertex parameter was introduced 
which results  in slightly different values for 
$T_N$ compared to the scheme used in the present paper.
\end{thebibliography}
\end{document}